# Ultrafast Anisotropic Polarization Dynamics of Electronic Ferroelectrics LuFe$_2$O$_4$

H. Yu[1], Y. Fukada[2], G. Nishida[2], K. Takubo[1], T. Ishikawa[1], S. Koshihara[1], N. Ikeda[2] and Y. Okimoto*[1]

[1] Department of Chemistry, Tokyo Institute of Technology, 2-12-1, Meguro, Tokyo 152-8551, Japan.,
[2] Department of Physics, Okayama Univ., 3-1-1, Tsushimanaka, Okayama, 700-8530, Japan.

(＊Electronic mail: okimoto.y.aa@m.titech.ac.jp)

**Abstract**

Femtosecond time-resolved second harmonic generation (SHG) measurements were performed at room temperature on a single crystal of LuFe$_2$O$_4$, which has recently attracted attention as an electronic ferroelectric material in which polarization occurs due to the real-space arrangement of Fe$^{2+}$ and Fe$^{3+}$. We irradiated the crystal with 800 nm laser pulses, traced the subsequent SHG changes, and observed anisotropic polarization dynamics: immediately after photoexcitation, the in-plane (*ab*-plane) component of the SHG greatly reduced, while the inter-plane (*c*-axis) SHG increased. This means that a hidden photoexcited state, in which the in-plane charge ordering is broken but the polarization along *c*-axis direction enhanced, has been created by photoexcitation, indicating that the electronic ferroelectricity can be controlled on a femtosecond scale.

**Introduction**

Ferroelectric materials have polarization (electric dipole moments) derived from their inversion symmetry-breaking crystal structures, whose direction can be controlled by electric fields. This makes ferroelectrics suitable for various industrial applications, such as memory devices. Therefore, the development of more energy-efficient and rapidly responsible ferroelectric materials is a critically important research topic in material science.

Recently, "electronic ferroelectrics" have garnered attention as a novel type of ferroelectric material [1-3]. Unlike traditional ferroelectrics, in which spontaneous polarization arises from ionic displacements or the ordering of molecules' orientations, electronic ferroelectrics exhibit spontaneous polarization due to the ordering of electrons, which breaks the inversion symmetry [4]. In electronic ferroelectrics, polarization can be altered by moving electrons, which are much lighter than ions, potentially allowing for the control of polarization at low electric fields [5] and at ultrafast speed. From this perspective, electronic ferroelectrics are anticipated as new materials that could surpass conventional ferroelectrics in performance and functions.

In 2005, Ikeda *et. al.* proposed the iron composite oxide $R$Fe$_2$O$_4$ (where $R$ is a trivalent rare-earth ion) as an example of electronic ferroelectrics [1]. This crystal consists of layers of $R$ ions and oxygen ions ($R$-layer) and a double layer of Fe ions and oxygen ions ($W$-layer) [6-8]. Notably, as shown in Fig. 1(a), in each $W$-layer, $Fe^{3+}$ (red circles) and $Fe^{2+}$ (blue circles) demonstrate a three-fold periodic charge order in the triangular lattice plane. Different numbers of $Fe^{2+}$ and $Fe^{3+}$ ions exist in the upper and lower layers of the $W$-layer. Consequently, the $W$-layer itself exhibits a dipole moment in the direction indicated by the arrows in Fig. 1(a), leading to spontaneous polarization throughout the crystal [1,5-9]

Another characteristic feature of this system is the issue of the correlation of the electric polarization along the $c$-axis. The total electric polarization of the $R$Fe$_2$O$_4$ crystal is formed by the sum of each polarization in $W$-layers as shown in Fig. 1(a), but all the directions of the polarizations in a crystal are not aligned along the $c$-axis. Fujiwara *et al.* demonstrated through temperature-dependent neutron scattering measurements of YbFe$_2$O$_4$ that the spin correlation length along the $c$-axis is approximately 50 nm at room temperature [10, 11]. From this, the length of polarization alignment along the $c$-axis in $R$Fe$_2$O$_4$ is also expected to be similar order. Figure 1(b) is a schematic diagram illustrating the $c$-axis correlation in $R$Fe$_2$O$_4$. The gray block represents the stacks of some polar $W$-layers aligned in the same direction. The thickness of these blocks along the $c$-axis corresponds to the correlation length of electric polarization. Thus, in $R$Fe$_2$O$_4$, not only the long-range three-fold periodic order of $Fe^{2+}$ and $Fe^{3+}$ in the $W$-layer but also the short-range correlations of the dipole moment of the $W$-layer along the *c*-axis, contribute to the macroscopic polarization of the system.

In this paper, we focus on LuFe$_2$O$_4$ among the $R$Fe$_2$O$_4$ series and attempt to control its polarization state by manipulating electrons in iron cations using femtosecond laser pulses. For studying the polarization dynamics of ferroelectrics, the method of time-resolved second-harmonic generation (SHG) measurements is useful [3, 12-19]. This method involves exciting the crystal with femtosecond pulses, and then measuring the SHG change with another laser pulse, which allows the polarization dynamics of the system to be traced on a femtosecond scale. This technique has been used for various ferroelectric materials to study the dynamics of ferroelectrics, control the polarization rapidly, and search for the so-called hidden state harbored in the ferroelectric materials [3, 16, 18].

In general, the SHG intensity ($I_{SH}$) is represented as $I_{SH} \propto |\boldsymbol{P}^{(2)}|^2 \propto |\varepsilon_0 \chi^{(2)} : \boldsymbol{EE}|^2$, using the second order nonlinear susceptibility tensor $\chi^{(2)}$. Here, $\varepsilon_0$ is the permittivity in vacuum, $\boldsymbol{E}$ is the electric field of the incident light, and $\boldsymbol{P}^{(2)}$ represents the second order nonlinear polarization. Our recent studies reported observations of SHG and its azimuth angle dependence in a single crystal [20] and thin film [21] of YbFe$_2$O$_4$, clarifying not only that the crystal has a polar structure but that the symmetry belongs to point group Cm of the monoclinic crystal. Figure 1(c) shows the azimuth angle dependence of $I_{SH}$ on the *ab*-plane of the LuFe$_2$O$_4$ crystal used in this study. Red circles represent $I_{SH}$ polarized in the *a*-axis direction ($I_{ab}{}^a$) and blue circles in the *b*-axis direction ($I_{ab}{}^b$). The polarization of incident electric field was rotated using a half-wave plate to the azimuth angle *θ*, which defined as the angle between the polarization of incident electric field and the *a*-axis. For both polarizations, a four-leaf clover-like azimuth angle dependence was observed, similar to YbFe$_2$O$_4$. Under the Cm point group symmetry, the reduced $\chi^{(2)}$ tensor is represented as follows [22].

$$\chi^{(2)} = \begin{pmatrix} d_{11} & d_{12} & d_{13} & 0 & d_{15} & 0 \\ 0 & 0 & 0 & d_{24} & 0 & d_{26} \\ d_{31} & d_{32} & d_{33} & 0 & d_{35} & 0 \end{pmatrix}. \quad (1)$$

Using this, when light is irradiated on the *ab*-plane, the azimuth angle dependence of the $I_{SH}$ radiated in the *a*-axis and *b*-axis polarization is described as

$$I_{ab}{}^a \propto (d_{11}\cos^2\theta + d_{12}\sin^2\theta)^2, \quad (2)$$

$$I_{ab}{}^b \propto (d_{26}\sin 2\theta)^2. \quad (3)$$

The solid lines in Fig. 1(c) represent the fitting results of observed $I_{ab}$ using equations (2) and (3), indicating that the LuFe$_2$O$_4$ crystal also has the symmetry belonging to point group Cm. On the basis of those results of SHG, in this study, we investigate the ferroelectric polarization dynamics and its anisotropy in LuFe$_2$O$_4$ through the time-resolved SHG measurements after optically disturbing the alignment of Fe ions and reveal not only a hidden photoexcited state created by light but try to photonically control the electronic ferroelectricity on a femtosecond scale at room temperature.

## Experimental

The LuFe$_2$O$_4$ single crystals were synthesized using the floating zone method: Fe$_2$O$_3$ and Lu$_2$O$_3$ powders were mixed in a prescribed ratio, shaped under hydrostatic pressure, and sintered in air for 10 hours. The sintered rod was then melted by a floating zone furnace to obtain a single crystal [10, 23]. Details about evaluating the crystallinity and the development of charge order of synthesized crystal was performed in supplementary material 2 [24]. The crystal orientation was determined using X-ray diffraction measurement, and the *ab* and *ac* planes were obtained by cutting with a diamond cutter. *a* and *c* axes polarized reflectivity spectra were measured using a Fourier transform interferometer and a grating spectrometer. Optical conductivity spectra were derived from reflectivity based on Kramers-Kronig analysis. Time-resolved SHG measurement was conducted with a regenerated amplified mode-locked Ti: Sapphire laser (wavelength ≈800 nm, pulse with ≈30 fs, repetition rate: 1 kHz) as a light source. (The experimental setup is shown in Fig. 2(a).) The laser pulse was split into two pulses; one of them is used for the pump light and directly irradiating the sample to resonant from *d-d* transitions in LuFe$_2$O$_4$. An optical parametric amplifier was used to transfer the wavelength of another pulse into 1300 nm for probe light. The polarization direction of the pump light was always set parallel to the *a*-axis, matching the polarization component of LuFe$_2$O$_4$. The polarization of the probe light (≈1300 nm) was controlled using a λ/2 plate before irradiating the sample. The incident angle of the probe light is estimated as about 10 degrees. The emitted SHG light (650 nm) from the sample polarized in a specific direction was detected by a photomultiplier tube after removing the fundamental wave of 1300 nm with a high-pass filter and a grating spectrometer.

## Results and discussions

To clarify the electronic structure of LuFe$_2$O$_4$ and its anisotropy, in Fig. 2(b), we show σ(ω) in LuFe$_2$O$_4$ polarized along *a*-axis (red line) and *c*-axis (blue line). In both polarization directions, there is a broad absorption peak at about 1.5 eV and a steep rise in absorption from around 3 eV. According to previous literatures [25, 26], the former is assigned to the Fe$^{2+}$→Fe$^{3+}$ *d-d* transition and the latter to the O 2*p*→Fe 3*d* charge transfer transition. In this *d-d* transition, the spectral weight in the *a*-axis polarized σ(ω) is higher than that in the *c*-axis polarization. In LuFe$_2$O$_4$, the distance between Fe ions in the *ab* plane is ≈0.346 nm, which is shorter than the inter-plane distance, ≈0.416 nm [27, 28]. This may be the reason for the difference in oscillator strength between in *a*-axis and *c*-axis polarized σ(ω). The downward red arrow in Figure 2(b) indicates the energy of the excitation light used for the time resolved SHG measurements. By the excitation causing the *d-d* transition, we can expect perturbing the charge ordering of the three-fold periodic Fe ions in the *W*-layer shown in Fig. 1(a) which realizes the electric polarization of LuFe$_2$O$_4$.

Figure 3(a) shows the time and excitation intensity dependence of the relative change in the intensity of the $a$-axis polarization component of the SHG light ($\Delta I_{ab}{}^{a}/I_{ab}{}^{a}$) after photoirradiation. Since $a$-axis polarized pulse ($\theta = 0°$) is used as the probe light, the SHG observed here is due to the $d_{11}$ component of Eq. (2). The time profile of $\Delta I_{ab}{}^{a}/I_{ab}{}^{a}$ shows a steep decrease immediately after light irradiation, then recovers slightly, but even after 200 ps $I_{ab}{}^{a}$ hardly returns to the value before photoexcitation. The magnitude of this change increases as the excitation light density is increased, and a decrease in SHG of ≈29% was observed at an excitation fluence of ≈10.7 mJ/cm$^2$.[29]

Figure 3(b) shows the azimuth angle dependence of $I_{SH}$ emitted in $a$ and $b$-axis polarization before (-0.5 ps) and immediately after (≈0 ps) photoexcitation. (The excitation intensity is ≈10.7 mJ/cm$^2$.) Just after photoexcitation, the clover-like angular profile shrank overall with maintaining its shape. This indicates that the magnitude of the electric polarization is reduced by photoexcitation, but the symmetry of the system (Cm) is conserved. The solid lines in the figure are the result of fitting based on Eqs. (2) and (3). The tensor components $d_{11}$, $d_{12}$, and $d_{26}$ are found to decrease by ≈14 % without changing their ratios compared to before excitation. It can be concluded that the charge ordering of Fe ions is disturbed by photoexcitation, resulting in a decrease in the $ab$ plane projection component of the electric polarization.

To quantitatively understand the time profile of $\Delta I_{ab}{}^{a}/I_{ab}{}^{a}$ in Fig. 3(a), we assume the following function when $t > 0$ as

$$F_{ab}(t) = A_0 - A_1 \exp\left(-\frac{t}{\tau_1}\right) - A_2 \exp\left(-\frac{t}{\tau_2}\right). \qquad (4)$$

Here, $\tau_1$ and $\tau_2$ are decay times and $A_0$, $A_1$, $A_2$ are constants to describe the time dependence. As shown by the solid lines in Fig. 3(a), we could well fit the observed time dependence by using Eq. (4) convoluted with the instrumental response function. The fitting result indicated that $\tau_1 \approx 0.12$ ps and $\tau_2 \approx 0.29$ ps at all the excitation intensity.

The next issue to discuss is the photoinduced dynamics exhibited by the $c$-axis polarization component of SHG. Figure 1(d) shows the azimuth angle dependence of $I_{SH}$ polarized along $a$-axis (red circles) and $c$-axis (blue circles) in the $ac$ plane of the LuFe$_2$O$_4$ crystal in the ground state; unlike the case for the $ab$ plane, a distorted angular profile is observed. Since this asymmetric profile could not be described by only a single domain with Cm symmetry, we postulated the existence of domains in which only the projection component of the polarization onto the $ab$ plane is inverted (60° domain), as shown in Supplement Material 1 [24]. This is caused by a 180° rotation of the phase of the $W$-layer stacking and is often observed in this system [30]. The angular profile of the SHG observed in the $ac$ plane is given by Eqs. (S1) and (S2) in Supplement Material 1 [24]. The SHG signal generated in this case is described by $d_{11}$ plus other tensor components ($d_{13}$, $d_{15}$, $d_{31}$, $d_{33}$, and $d_{35}$). The solid line in Fig. 1(d) is the result of fitting using Eqs. (S1) and (S2), confirming that the symmetry of the LuFe$_2$O$_4$ crystal can be described by Cm, even when measured on the *ac* plane.

Figure 4(a) shows the time dependence of the relative change in *c*-axis polarized SHG pulse ($\Delta I_{ac}^{c}/I_{ac}^{c}$) after photoexcitation in the *ac* plane of the LuFe$_2$O$_4$ crystal. (The polarization angle of the incident probe light is $\theta = 110°$, the direction in which the *c*-axis polarization component of the generated SHG light is strongest.) Immediately after photoexcitation, $\Delta I_{ac}^{c}/I_{ac}^{c}$ decreases instantaneously as in the *ab* plane case (Fig. 3(a)), but then it is observed to increase after ≈0.2 ps. Furthermore, increased $I_{SH}$ slowly returns to the pre-excitation state. This result is qualitatively different from the dynamics seen in the time profile of $\Delta I_{ab}^{a}/I_{ab}^{a}$ in Fig. 3(a), where suppression of the SHG survives long after photoexcitation, and indicates that the *c*-axis polarization component increases after photoexcitation and slowly returns to the original state. Such anisotropic polarization dynamics is a characteristic feature of LuFe$_2$O$_4$ crystal.

To further quantify the dynamics in the *ac* plane, we assume the following function $F_{ac}(t)$ when $t > 0$ to reproduce the time evolution of the SHG after photoexcitation:

$$F_{ac}(t) = B_0 - B_1 \exp\left(-\frac{t}{\tau_1}\right) - B_2 \exp\left(-\frac{t}{\tau_2}\right) - B_3 \exp\left(-\frac{t}{\tau_3}\right). \quad (5)$$

Here, $\tau_1$ and $\tau_2$ are the relaxation times and $\tau_3$, $B_0$, $B_1$, $B_2$, and $B_3$ are fitting parameters to describe each amplitude. We performed the fitting analysis by convoluting the SHG time profiles with the instrumental response function, as we did in the *ab* plane. As a result, good fits were obtained for all excitation fluences as represented by the solid line in Fig. 4(a). The magnitude of $\tau_3$ is almost the same for all excitation densities and can be estimated as $\tau_3 \approx 4.5$ ps.

The unique behavior of the SHG dynamics in the *ac* plane indicates that variations in tensor components other than $d_{11}$ and $d_{12}$ also play an important role in the polarization dynamics of the system. Figures 4(b) and 4(c) show the azimuth angle dependence of the SHG polarized along *a*-axis (b) and *c*-axis (c), which were measured before, just after, and 0.2 ps after the photoirradiation. (The filled triangles in Fig. 4(a) denote the respective delay times.) At all delay times, the azimuth angle dependence of SHG retains the asymmetric shape seen before photoexcitation. However, the magnitude of the overall angular profile decreases immediately after optical excitation and then increases again after 0.2 ps. We performed fitting analysis of this azimuth angle dependence in the *ac* plane based on Eqs. (S1) and (S2) and calculated the $\chi^{(2)}$ tensor components at each delay time. The solid black lines in Figs. 4(b) and 4(c) show the results of the fitting analysis. Figure 5 shows the time dependence of the relative rates of change of $d_{11}$, $d_{35}$, and $d_{33}$ obtained from the fitting analysis. It shows that $d_{11}$ and $d_{35}$ decrease immediately after excitation and then recover slightly but remain negative in sign, whereas $d_{33}$ slightly decreases immediately after irradiation and then increases by ≈3% more than before photoexcitation. Since $d_{11}$ and $d_{33}$ are proportional to the degree of inversion symmetry breaking in the *a*- and *c*-axis directions, respectively, it can be seen that at ≈0.2 ps after photoexcitation, the in-plane component of polarization is reduced, but the *c*-axis component of polarization is rather enhanced compared to before the excitation.

Based on these results, we discuss the polarization dynamics of $LuFe_2O_4$. Figure 6 shows a schematic diagram of the change of state of the *W*-layer and the Fe ions in it upon photoexcitation. The in-plane *d-d* transition of $Fe^{2+} \rightarrow Fe^{3+}$ by photoexcitation leads to the Frank-Condon (FC) state in which the threefold periodic order of $Fe^{2+}$ and $Fe^{3+}$ in the ground state is disturbed [see Fig. 1(a)]. Via the FC state, photoexcited state appears: First, the in-plane SHG signal of $LuFe_2O_4$ decreases significantly with the time constant of $\tau_1$ (≈0.12 ps), while not in the inter-plane $I_{SH}$, indicating that the *W*-layer is composed of a $Fe^{2+}$-rich layer (indicated by thin red circles) and a $Fe^{3+}$-rich layer (indicated by thin blue circles) whose three-fold periodic order remains disturbed. In this state, the direction of polarization is on average oriented toward the *c*-axis, which explains that $d_{11}$ decreases but $d_{33}$ scarcely changes. The right panel of Fig. 6(b) shows a schematic of the correlation of the *c*-axis polarization described in Fig. 1(b). In this time region, the gray blocks are composed of the charge disproportionated *W*-layers and stacked in the *c*-axis direction, keeping the correlation length seen in the ground state (≈50 nm).

What should be noted is the dynamics represented by the time constant $\tau_2$ (≈0.29 ps) [Fig. 6(c)]. In this time region, the $I_{SH}$ in *ab* plane slightly rises due to the relaxation of some small region of the excited area [18], but most of the photoexcited region remains reduced and the *W*-layers still lose their in-plane three-fold periodic order as shown in Fig. 6(b). By contrast, the inter-plane SHG shows an instantaneous increase. The reason for this is not yet clear, but one possibility is that the correlation length of the electric polarization has increased as shown in the right schematics of Fig. 6(c). In the *c*-axis polarized SHG radiation, if the volume of one polarization domains increases, total $I_{SH}$ also increases because of the reduction of interference. In the present situation, the *W*-layers are composed of $Fe^{2+}$-rich and $Fe^{3+}$-rich layers without in-plane long-range order, and it is possible that the total electric polarization direction is more aligned along the *c*-axis to gain long-range Coulomb energy between the *W*-layers and that the correlation length along the *c*-axis increases. According to the scenario, the variation of $d_{33}$ in Fig.5 results from the reorientation of the stack of the $Fe^{2+}$-rich and $Fe^{3+}$-rich layers, suggesting that the *c*-axis correlation increases by about 3% at 0.2 ps.

This increased SHG slowly recovers with a time constant of $\tau_3$ (≈4.5 ps), and the correlation length of the *c*-axis polarization returns to the original state [Fig. 6(d), right]. Thus, not only the charge ordering state of the Fe ions but also the changes in the *c*-axis correlation between *W*-layers are intertwined and determine the ultrafast anisotropic polarization dynamics of $LuFe_2O_4$.

Another possible scenario for the origin of the observed instantaneous increase in the *c*-axis polarized SHG in $LuFe_2O_4$ is a change of the crystal structure itself with photoexcitation. In the previous section, we considered a scenario in which the crystal hardly changes and the ordered state of Fe ions varies. However, if the position of the ions constituting the crystal moves upon photoexcitation, the magnitude of the SHG as well as $\chi^{(2)}$ tensor components of the system changes as a result. For example, if the distance between the *W*-layers increases after light irradiation, the *c*-axis magnitude of the electric polarization shown in Fig. 1(a) can increase, and the magnitude of the $d_{33}$ component of the $\chi^{(2)}$ tensor should increase accordingly. To investigate this experimentally, in addition to first-principles calculations, time-resolved crystallographic studies on the 100 fs scale using femtosecond X-ray and electron beam pulses, which have been actively pursued in recent years [13, 31], are needed, and this would deserve future research.

## Summary

Optical conductivity, SHG azimuth angle dependence, and anisotropy of femtosecond time-resolved pump-probe SHG changes were measured for $LuFe_2O_4$ single crystal which is noticed as an electronic ferroelectric material. The optical conductivity spectrum showed anisotropy consistent with charge order alignment. SHG azimuth angle dependence on *ab* plane and *ac* plane confirmed $LuFe_2O_4$ has monoclinic Cm symmetry and determined the ratio of the $\chi^{(2)}$ tensor components. Time-resolved SHG measurements revealed anisotropic dynamics: 800 nm pulse irradiation causing $Fe^{2+} \rightarrow Fe^{3+}$ transition that disturbs iron ion charge order reduces $I_{SH}$ emitted from *ab* plane, while rather increases $I_{SH}$ emitted from *ac* plane, indicating enhancement of *c*-axis component of the electric polarization by light. These anisotropic dynamics was discussed in terms of charge order and disorder, and variations of correlation length of polarization between *W*-layers. This study demonstrates the ultrafast and room-temperature control of electric polarization induced by electronic order by light irradiation, suggesting possibilities for applications in high-speed optical communication and ferroelectric memory devices.

## Acknowledgments

The authors thank R. Seimiya, R. Ota, S. Ogasawara, K. Fujiwara and T. Fujii for their technical assistance and discussion. This research was supported by Japan Society for the Promotion of Science (JSPS) KAKENHI Grants Nos. JP18H05208, JP18H02057, JP19H01827, JP22H01153, JP22H01942, JP22H01149, JP20H05147.

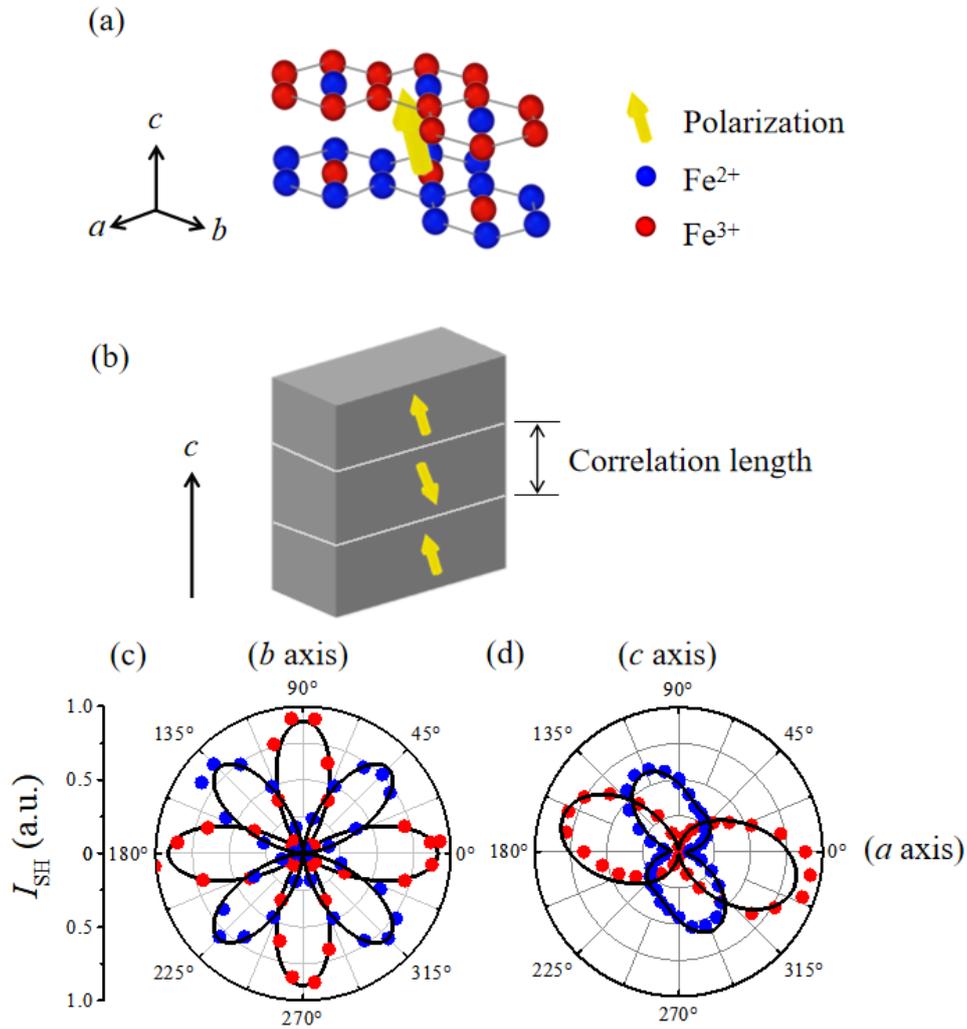

Fig. 1 (a): Schematic diagram of $W$-layer in $LuFe_2O_4$ crystal. The red and blue circles represent $Fe^{3+}$ and $Fe^{2+}$, respectively, and the yellow arrow indicates the direction of polarization.

(b): Schematic diagram showing how $W$-layers are stacked along the $c$-axis direction in a $LuFe_2O_4$ crystal. The gray block indicates a stack of several $W$-layers with aligned polarization directions, and the thickness (correlation length) is estimated to be approximately 50 nm.[10]

(c): Azimuth angle dependence of $a$-axis polarized (red circle) and $b$-axis polarized SHG (red circle) in $ab$ plane.

(d): Azimuth angle dependence of $a$-axis polarized (red circle) and $c$-axis polarized SHG (red circle) in $ac$ plane. The solid lines are the fitting result assuming monoclinic Cm.

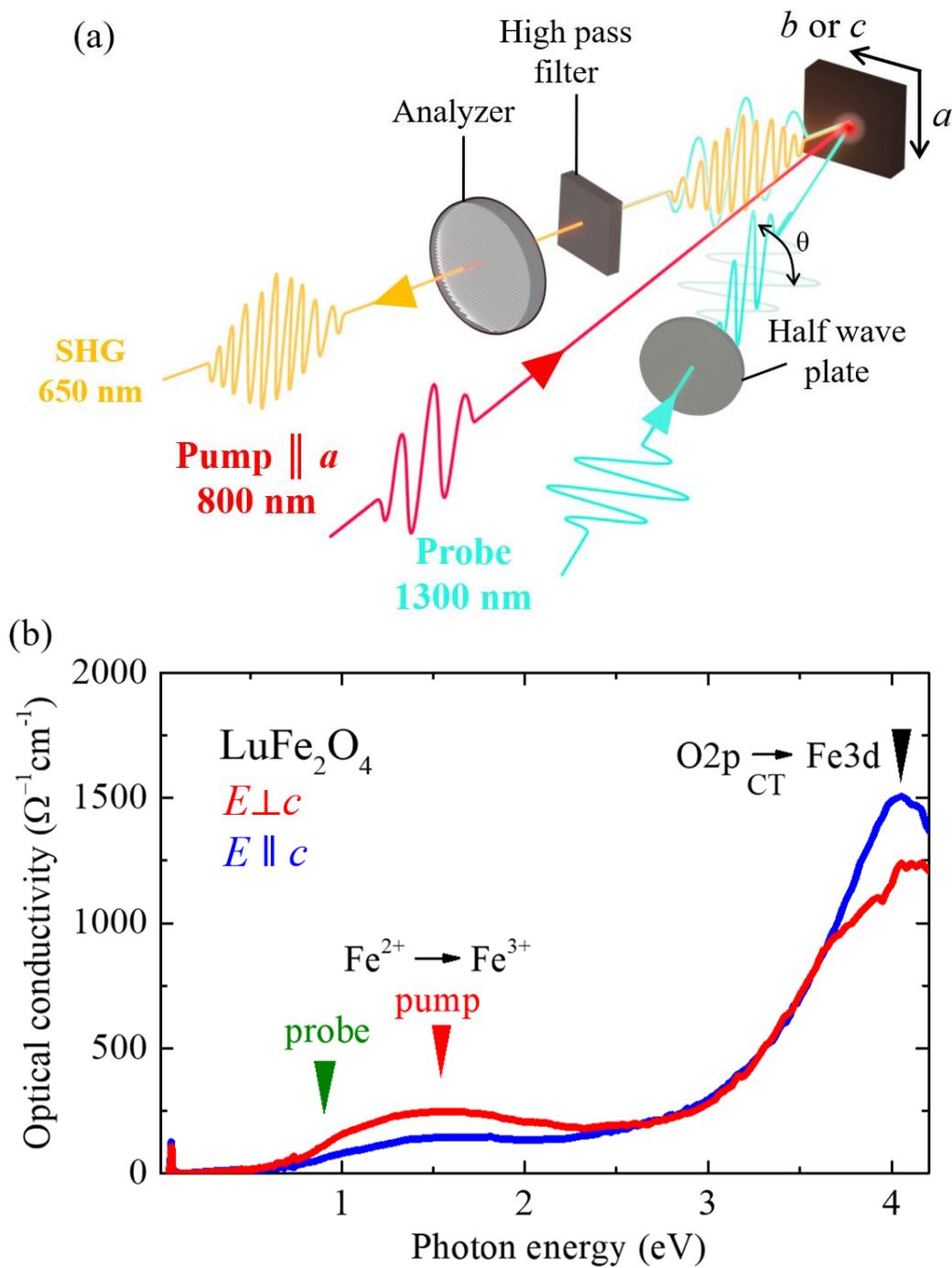

Fig. 2 (a): Schematic diagram of the experimental setup for measuring time-resolved SHG changes.

(b): Optical conductivity spectra in $LuFe_2O_4$ crystal. The red line is the spectrum of $E\perp c$, and the blue line is the spectrum of $E\|c$.

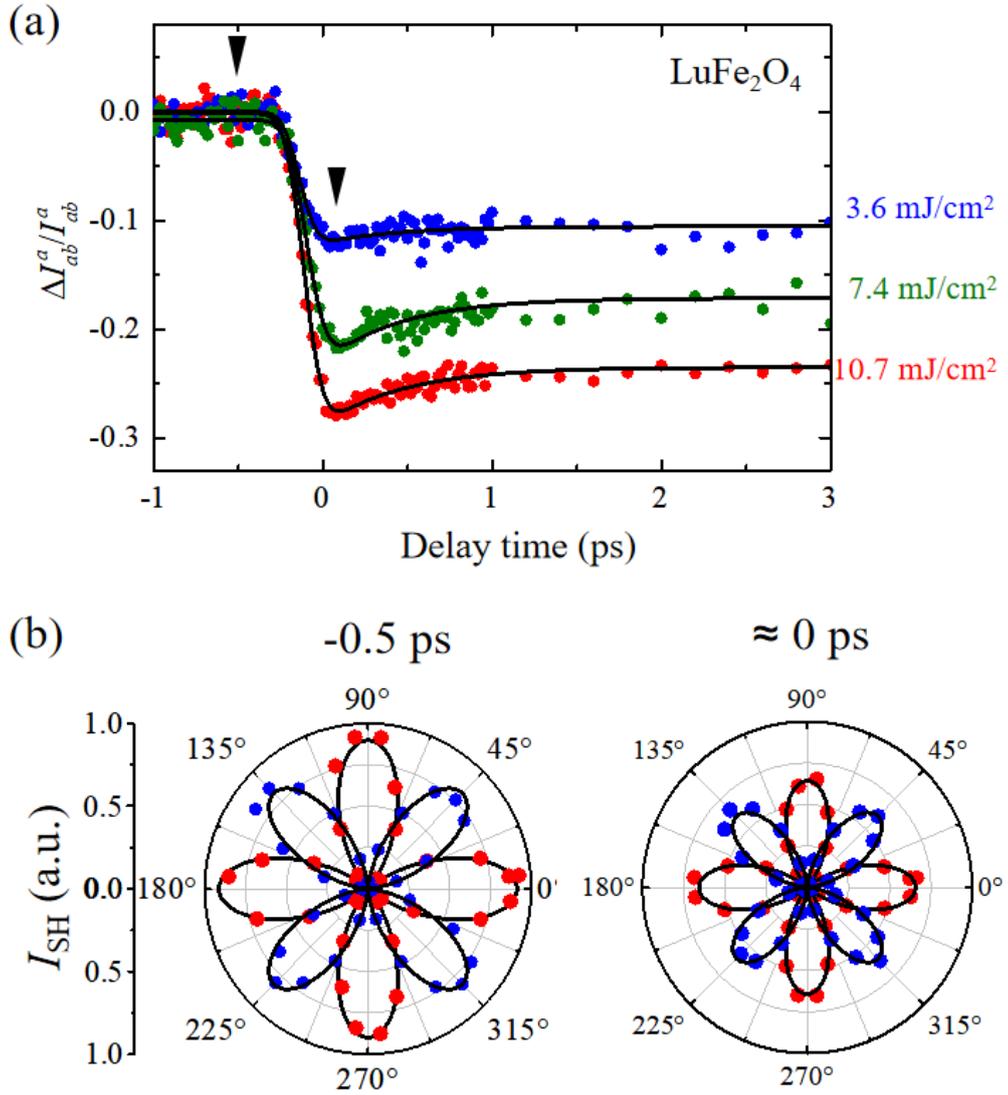

Fig. 3(a): Time evolutions of *a*-axis polarized SHG intensity after photoexcitation of selected fluence on the *ab* plane of LuFe$_2$O$_4$ crystal. The solid line is the fitting result based on Eq. (4) (see text).

(b): Azimuth angle dependence of SHG of the *a*-axis (red circle) and *b*-axis polarization (blue circle) polarization before and immediately after photoexcitation.

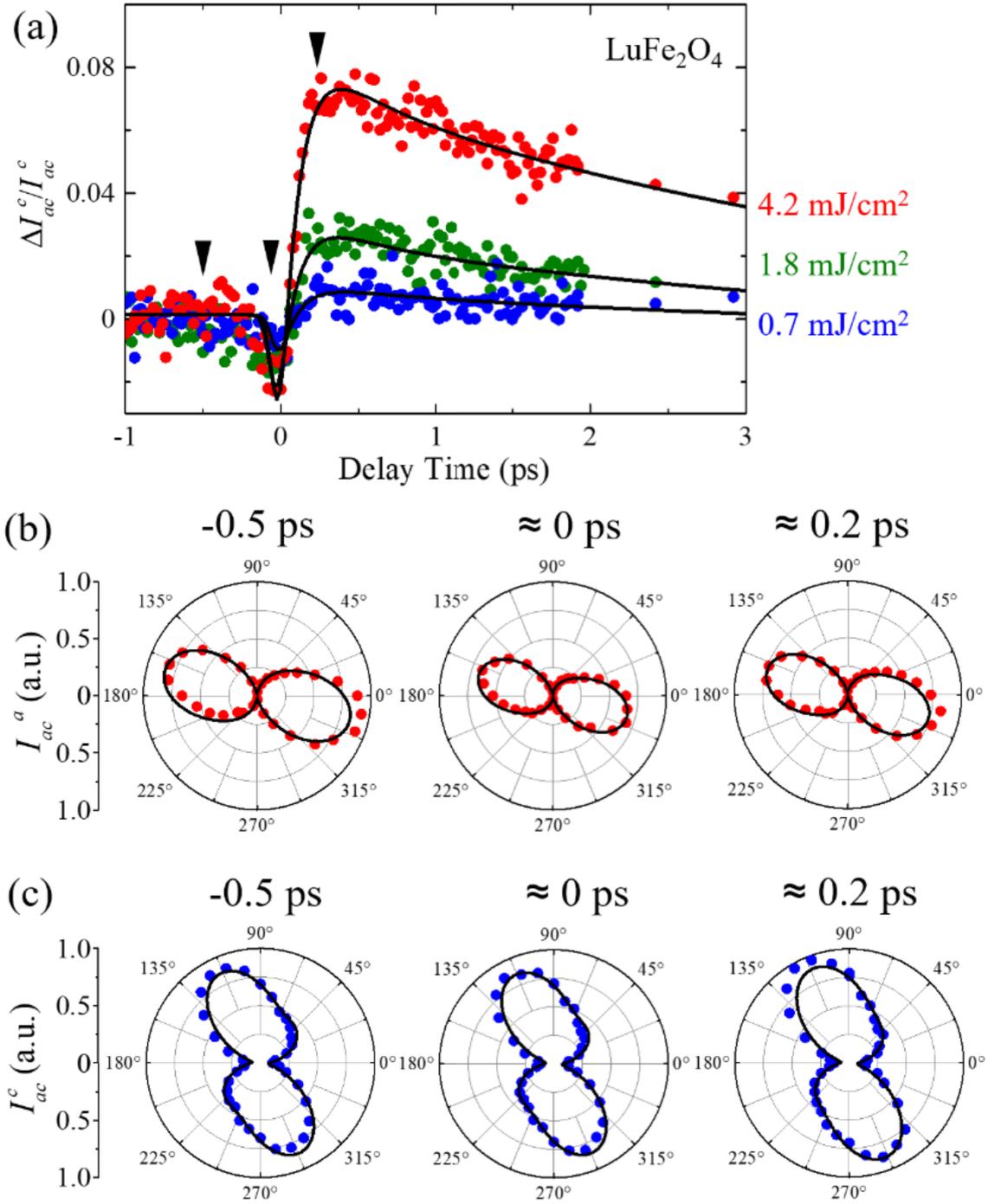

Fig. 4(a): Time evolutions of *c*-axis polarized SHG intensity after photoexcitation of selected fluence on the *ac* plane of LuFe$_2$O$_4$ crystal. The solid line is the fitting result based on Eq. (5) (see text).

(b) and (c): Azimuth angle dependence of *a*-axis [(b)] and *c*-axis polarized [(c)] SHG at -0.5 ps, 0 ps, and 0.2 ps.

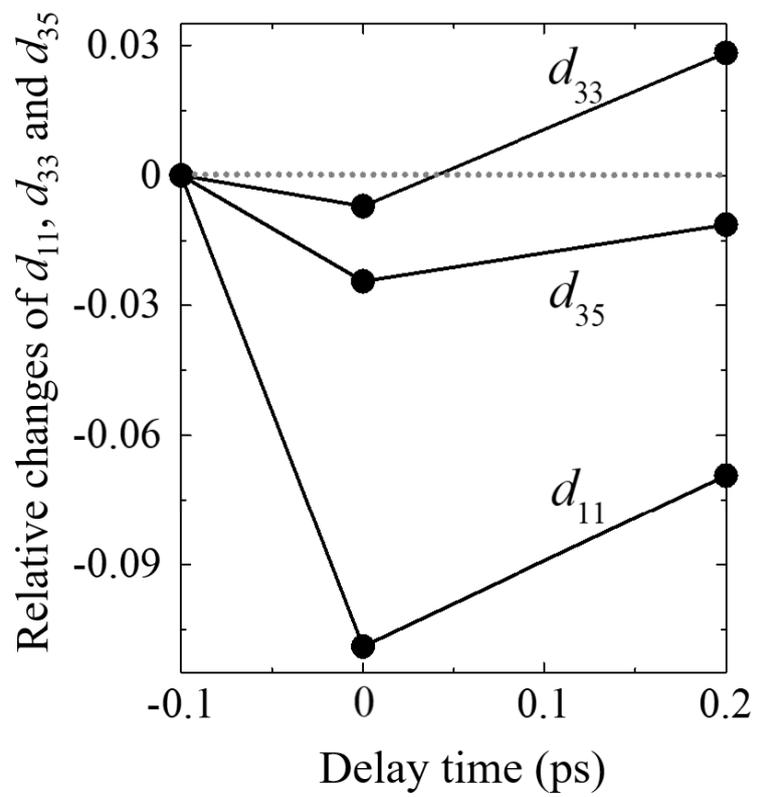

Fig. 5: Time evolution of the selected reduced $\chi^{(2)}$ tensor components ($d_{11}$, $d_{35}$, $d_{33}$) after photoexcitation.

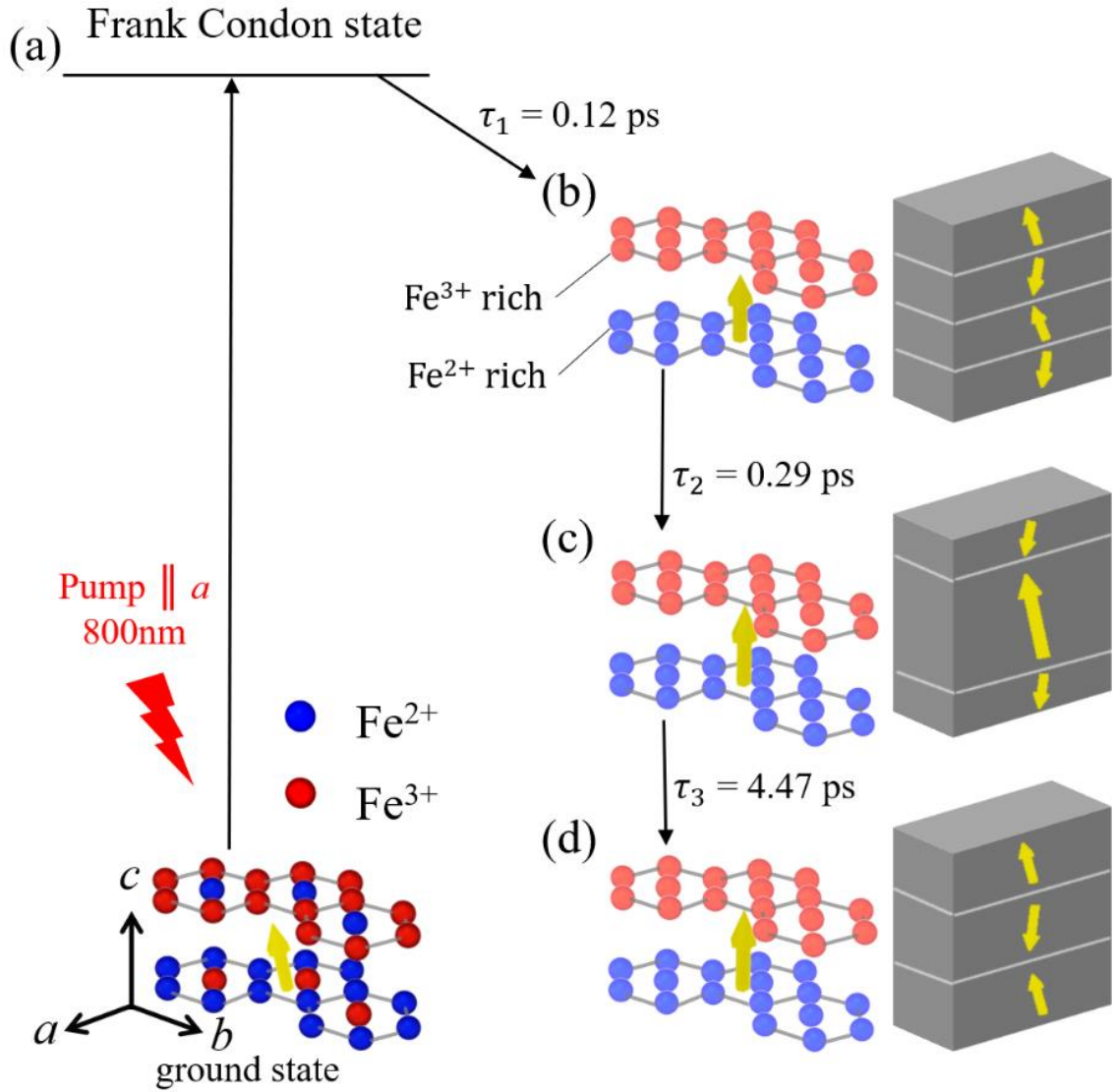

Fig. 6: Schematic diagram of the dynamics of iron ions in the *W*-layer after photoexcitation in LuFe$_2$O$_4$. By irradiating 800 nm pulse, the ground state in which Fe$^{2+}$ and Fe$^{3+}$ is disturbed and Franck-Condon state occurs (a). After that, the *W*-layer becomes composed of an average Fe$^{2+}$-rich layer (light red) and Fe$^{3+}$-rich layer (light blue) and stacking of *W*-layers successively varies [(b)-(d)]. The gray block in the right column represents the *W*-layer stacking described in Figure 1(b). The yellow arrow indicates the polarization direction.